\documentclass[twocolumn,english,prl]{revtex4-1}
\usepackage[T1]{fontenc}
\usepackage{amstext}

\makeatletter

\@ifundefined{textcolor}{}
{%
 \definecolor{BLACK}{gray}{0}
 \definecolor{WHITE}{gray}{1}
 \definecolor{RED}{rgb}{1,0,0}
 \definecolor{GREEN}{rgb}{0,1,0}
 \definecolor{BLUE}{rgb}{0,0,1}
 \definecolor{CYAN}{cmyk}{1,0,0,0}
 \definecolor{MAGENTA}{cmyk}{0,1,0,0}
 \definecolor{YELLOW}{cmyk}{0,0,1,0}
}

\makeatother

\usepackage{babel}
\begin{document}
\global\long\def\ket#1{\left|#1\right\rangle }

\global\long\def\bra#1{\left\langle #1\right|}

\global\long\def\braket#1#2{\left\langle #1\left|#2\right.\right\rangle }

\global\long\def\ketbra#1#2{\left|#1\right\rangle \left\langle #2\right|}

\global\long\def\braOket#1#2#3{\left\langle #1\left|#2\right|#3\right\rangle }

\global\long\def\mc#1{\mathcal{#1}}

\global\long\def\nrm#1{\left\Vert #1\right\Vert }

\title{\textmd{\normalsize{}Universal features in the efficiency of ultra
hot quantum Otto engines}}

\author{Raam Uzdin}

\author{Ronnie Kosloff }

\address{Fritz Haber Research Center for Molecular Dynamics, Hebrew University
of Jerusalem, Jerusalem 91904, Israel}

\email{raam@mail.huji.ac.il}

\email{ronnie@fh.huji.ac.il}

\begin{abstract}
We study ``internal'' work optimization over the energy levels of
a generic hot quantum Otto engine. We find universal features in the
efficiency that resembles the classical ``external'' power optimization
over the coupling times to the thermal baths. It is shown that in
the ultra hot regime the efficiency is determined solely by the optimization
constraint, and independent of the engine details. We show that for
some constraints the radius of convergence of the perturbative approach
used in the classical analysis is zero even for very arbitrarily low
efficiencies at small temperature difference.
\end{abstract}
\maketitle
Carnot's discovery of a universal efficiency upper bound for heat
engines had a profound impact on physics and engineering. Yet, in
practice to approach this efficiency bound the system needs to be
reversible and that leads to infinitely slow cycle time and vanishing
power output. This motivated extensive studies under the title of
``finite-time thermodynamics'' (see \cite{salamon01,AndresenFiniteTimeThermo2011}
for review articles). More importantly, efficiency is only a secondary
design goal. First the engine must be capable of doing the task it
is designed to: lifting a weight in a given time, accelerating a car
etc. In general the efficiency depends on the heat transfer mechanism
between the system and the bath. Nonetheless, some universal features
were discovered when the power output is maximal. In this work we
study the universality of efficiency at maximal output of quantum
Otto engines. In the engines studied here, the working substance is
a single particle that constitutes an $N$-level system. In the adiabatic
stroke of the quantum Otto engine the levels of the particle must
be varied in time. In real systems this level variability is limited
by practical considerations. For example in Zeeman splitting the maximal
gap is determined by the maximal available external magnetic field.
In other systems it may be the power of the laser. In this work we
study the optimal output of engines subjected to this type of constraints.
We find that the details of the engine are irrelevant when the baths
are very hot. The efficiency at maximal output is determined only
by the nature of the constraint and the temperatures. For some family
of constraints the universality features can be expressed using a
perturbative approach analogous to the classical analysis, but we
also identify constraints cannot be treated with perturbation theory.

Typically in classical engines the equation of state of the working
substance is known and the output optimization is done by changing
the coupling time to the baths. The output power may change from system
to system but it was observed that for some classes of classical engines
the efficiency at maximum power has universal features. In particular
in \cite{espositoEfficiencyLowDissipation,VanDenBroeckEffMaxLowDiss13}
it was shown that in the low dissipation limit the efficiency at maximum
power satisfies:
\begin{equation}
\frac{\eta_{c}}{2}\le\eta_{LD}\le\frac{\eta_{c}}{2-\eta_{c}},\label{eq: eta LD bound}
\end{equation}
where $\eta_{c}$ is the Carnot efficiency. The same results were
obtained in \cite{WangEffMaxPower12,ChenEffMaxPower89} for different
thermalization mechanisms. In the low dissipation scenario, the special
case where the coupling coefficients to the cold and hot baths is
the same (symmetric case) yields the Curzon-Ahlborn \cite{ChambadalCA,novikov1958efficiency,curzon75}
efficiency,
\begin{equation}
\eta_{CA}=1-\sqrt{T_{c}/T_{h}}.\label{eq: eta CA}
\end{equation}
$\eta_{CA}$ was originally obtained by applying the Newton heat transfer
law. Features of universality appear in the Taylor expansion of the
efficiency in terms of the Carnot efficiency. In \cite{EspositoRightLeftSym}
it was shown that:
\begin{equation}
\eta_{Pmax}=\frac{\eta_{c}}{2}+\frac{\eta_{c}^{2}}{8}+O(\eta_{c}^{3}).
\end{equation}
The $\frac{1}{2}$ factor of the linear term is universal and the
second term is universal for systems that have a ``left-right''
symmetry. For studies of efficiency in different model see \cite{allahmahler08,SeifetBrownianEffMaxPower,zhou2010minimal,allahaverdian2013}
and references therein.

In this work we study work optimization for quantum Otto engines \cite{SwapMultilevelArxiv}.
The working substance is a single $N$-level particle that is coupled
periodically to hot and cold baths. The properties of this quantum
working substance are determined by the level structure of the particle
when it is coupled to the hot and to the cold baths. We optimize the
level structure to produce maximal work output per cycle. When the
cycle time is fixed this is equivalent to power optimization. The
classical optimization described earlier can be called ``external''
as it involves the optimization of the coupling process to the external
bath. The maximum power in this case originates from the fact that
reaching a thermal equilibrium with the baths is time consuming. Yet
our optimization is ``internal'' as we optimize over the working
medium properties (the level structure). We will assume that the baths
is coupled for sufficiently long period to effectively reach equilibrium
for all practical purposes. This assumption is very reasonable when
only one particle needs to be thermalized (and not a whole medium
filled with particles). Yet, the analysis here includes the case of
partial ``swap'' thermalization \cite{SwapMultilevelArxiv} ($\xi\neq1$).
It is remarkable that for the most basic constraints the results of
this internal structure quantum optimization are identical to the
classical external coupling optimization. 

We consider a generic single particle four-stroke Otto cycle. In the
adiabatic strokes the energy levels of the particle (engine) change
in time without changing the populations (see discussion in \cite{SwapMultilevelArxiv}
on ways of achieving this in a short time). In the thermal strokes
the system is coupled to hot and cold baths. If the system is coupled
for periods that exceed a few relaxation times, it is plausible to
assume a full thermalization has taken place. As we show in this work,
some universal engine-independent features appear in the ultra-hot
limit where only the leading order in $\beta$ (inverse of the temperature)
is kept. In some cases, our results hold to order $\beta^{2}$ as
well.

The work output of an $N$-level ultra-hot swap engine is \cite{SwapMultilevelArxiv}:
\begin{equation}
W_{hot}^{ultra}=\frac{\xi}{2-\xi}\frac{1}{N}[(\beta_{c}+\beta_{h})\mc E_{c}\cdot\mc E_{h}-\beta_{c}\left|\mc E_{c}\right|^{2}-\beta_{h}\left|\mc E_{h}\right|^{2}],\label{eq: W}
\end{equation}
where $\mc E_{c(h),i}$ is the $i$-th cold (hot) energy level of
the engine. The energy levels are shifted so that $Mean(\mc E)=0$.
The work is energy shift invariant but a zero mean lead to a more
compact form. The swap parameter $0\le\xi\le1$ determines the degree
of thermalization in the thermal strokes of the engine. When $\xi=1$
a full thermalization takes place. This case should hold for any interaction
that lead to a practically full thermalization regardless of the mechanism
that generates it. Before proceeding we note that the norm of the
levels in the ultra hot regime is directly related to several key
quantities. For example, the internal energy when couple to the one
bath is $tr(\rho_{b}\hat{\mc E}_{b})=\frac{1}{N}\beta_{b}\left|\mc E_{b}\right|^{2}$,
the purity is $tr(\rho_{b}^{2})=\frac{1}{N}+\frac{1}{N^{2}}\beta_{b}^{2}\left|\mc E_{b}\right|^{2}$,
and the heat capacity is $C_{v}=\frac{1}{N}\beta_{b}^{2}\left|\mc E_{b}\right|^{2}$.
Another example for the norm significance will be given later on.
In \cite{SwapMultilevelArxiv} it was shown that once the variance
(or norm) of the hot and cold levels are fixed, the maximum work is
obtained when the energy vectors are parallel:
\begin{eqnarray}
\mc E_{c} & = & (1-\chi)\mc E_{h},\label{eq: x parallel}\\
0 & \le & \chi\le\eta_{c},\label{eq: engine cond}
\end{eqnarray}
where $\chi$, the compression deviation, is related to the compression
ratio via $\mc C=\frac{1}{1-\chi}$. Condition (\ref{eq: engine cond})
follows from the necessary condition for engine operation in the ultra-hot
regime $T_{c}/T_{h}\le\left|\mc E_{c}\right|/\left|\mc E_{h}\right|\le1$
(see \cite{SwapMultilevelArxiv}). The exact expression for the efficiency
of an Otto engine with uniform compression (\ref{eq: x parallel})
is \cite{SwapMultilevelArxiv}: 
\begin{equation}
\eta=1-\left|\mc E_{c}\right|/\left|\mc E_{h}\right|=\chi.\label{eq: eta x y}
\end{equation}
Despite the equality of (\ref{eq: eta x y}) it is useful to separate
the notation in order to prevent confusion. The maximal work in terms
of $\chi$ and the Carnot efficiency is:  
\begin{equation}
W_{\chi}=\frac{\xi}{2-\xi}\beta_{c}\chi(\eta_{c}-\chi)\frac{\left|\mc E_{h}\right|^{2}}{N},\label{eq: W max}
\end{equation}
where the subscript $\chi$ indicates that we have already imposed
the necessary but not sufficient optimality condition (\ref{eq: x parallel}).
Notice that $W_{\mc{\chi}}(\chi=0)=0$ (no compression) and $W_{\chi}(\chi=\eta_{c})=0$
(reversible limit in Otto engines). Since $\left|\mc E_{h}\right|\neq0$,
it follows from (\ref{eq: W max}) that a maximum exists in the domain
$\chi=\eta\in(0,\eta_{c})$. The maximal work in the ultra-hot regime
has an inherent universality. It depends only on the norms $\left|\mc E_{h}\right|^{2}$
and $\left|\mc E_{c}\right|^{2}$ (or $\left|\mc E_{h}\right|^{2}$
and the compression ratio). The specific energy levels structure plays
no role. All quantum Otto engines %
\footnote{Otto engines that reach full thermalization in the thermal stroke.
In addition we assume that the non-adiabatic losses in the adiabatic
strokes are eliminated using one of the methods described in \cite{SwapMultilevelArxiv}.%
} with the same energy variance $\left|\mc E_{h}\right|^{2}/N,\left|\mc E_{c}\right|^{2}/N$
will have the same efficiency and same maximal work per cycle (up
to the $\xi/(2-\xi)$ factor in (\ref{eq: W max})). The finer details
of the engine manifest themselves only at colder temperatures.

\section*{Work per cycle optimization}

We start with a few important cases that exemplify the kinship to
the classical case with very little algebra. First we choose the constraint
$\left|\mc E_{h}\right|=\text{const}$. Applying $\frac{d}{d\chi}W_{\chi}=0$
to (\ref{eq: W max}) with fixed $\left|\mc E_{h}\right|$ yields:
\begin{equation}
\eta_{\left|\mc E_{h}\right|}=\frac{\eta_{c}}{2},
\end{equation}
which is the lower limit on the efficiency in the low dissipation
model (\ref{eq: eta LD bound}). On the other hand the opposite constraint
$\left|\mc E_{c}\right|=\text{const}$ ($\left|\mc E_{h}\right|=\text{const}/(1-\chi)$)
yields:
\begin{equation}
\eta_{\left|\mc E_{c}\right|}=\frac{\eta_{c}}{2-\eta_{c}},
\end{equation}
which is the upper limit on the efficiency in the low dissipation
model (\ref{eq: eta LD bound}). When applying the symmetric constraint
$\left|\mc E_{c}\right|\left|\mc E_{h}\right|=\text{const}$ then:
\begin{equation}
\eta_{\left|\mc E_{c}\right|\left|\mc E_{h}\right|}=\eta_{CA}=1-\sqrt{1-\eta_{c}}.
\end{equation}
Although this specific symmetric constraint yields CA efficiency (\ref{eq: eta CA}),
we shall see that symmetry does not necessarily lead to the CA efficiency
in quantum Otto engines. The CA efficiency was observed in a specific
hot quantum engine in \cite{GevaCarnotTLS92}. Another important example
follows from the constraint $\alpha\left|\mc E_{c}\right|+(1-\alpha)\left|\mc E_{h}\right|=\text{const}$
that yields the maximum power efficiency:
\begin{equation}
\eta_{\alpha}=\frac{\eta_{c}}{2-\alpha\eta_{c}}.\label{eq: eta alpha}
\end{equation}
This efficiency form frequently appears in various classical systems
such as Brownian engines \cite{SeifetBrownianEffMaxPower}, system
operating in the low dissipation limit \cite{VanDenBroeckEffMaxLowDiss13},
and system with other thermalization processes \cite{WangEffMaxPower12,ChenEffMaxPower89}. 

The simple linear constraints studied above can be solved in a closed
form. In what follows we explore the low efficiency limit for a general
constraint and find universal features.

\section*{A general optimization constraint}

As an example for a non-trivial physical constraint that is characterized
by the energy norms, consider the quantum Otto engine studied in \cite{2spinDemagnetizationEngine,Optimal2spinEngine}.
This engine has four energy levels and it is comprised of two interacting
spins and an external time-dependent magnetic field. In order to have
the same population in the beginning and at the end of the adiabatic
evolution strokes a certain protocol must be applied. Using the optimal
protocol in \cite{Optimal2spinEngine}, the minimal time for the adiabatic
step is proportional to $\frac{1}{\left|\mc E_{h}\right|}+\frac{1}{\left|\mc E_{c}\right|}$
(to simplify (24) in \cite{Optimal2spinEngine} we considered the
limit $\omega_{f,}\omega_{i}\ll j$). Thus, for the engine to operate
at the minimal possible time (e.g to maximize the power) the constraint
is $\frac{1}{\left|\mc E_{h}\right|}+\frac{1}{\left|\mc E_{c}\right|}=\text{const}$.
This example shows that time optimization for maximal power by eliminating
the quantum non-adiabatic effects, manifest itself as an energy norm
constraint. In addition it clarifies that for observing universality
there is a justified need for a framework valid for more complicated
constraints. Applying $\frac{d}{d\chi}W_{\chi}=0$ to (\ref{eq: W max})
we get: 

\begin{equation}
\frac{\frac{d}{d\chi}\left|\mc E_{h}\right|}{\left|\mc E_{h}\right|}=\frac{(\eta_{c}-2\chi)}{2(\chi\eta_{c}-\chi^{2})}.\label{eq: gen dW 0}
\end{equation}
At this point we introduce the constraint function:

\begin{equation}
G(\left|\mc E_{c}\right|,\left|\mc E_{h}\right|)=\text{const}.\label{eq: G def}
\end{equation}
that can describe either an implementation constraint or a design
goal. writing: 
\begin{equation}
G((1-\chi)\left|\mc E_{h}(\chi)\right|,\left|\mc E_{h}(\chi)\right|)=\text{const}.\label{eq: G(chi,Eh)}
\end{equation}
we get the extra equation needed to find $\chi$. The only limitation
on $G$ is that (\ref{eq: G(chi,Eh)}) must provide a positive continuous
solution for $\left|\mc E_{h}\right|$ in the domain $0<\chi<\eta_{c}$.
When $\left|\mc E_{h}(\chi)\right|$ can be solved explicitly from
(\ref{eq: G(chi,Eh)}), then it can be used to evaluate the left hand
side of (\ref{eq: gen dW 0}) and obtain an explicit equation for
the optimal $\chi$. Yet, it is simpler to take the derivative of
(\ref{eq: G(chi,Eh)}), evaluate $\frac{d}{d\chi}\left|\mc E_{h}\right|/\left|\mc E_{h}\right|$
and then use it in (\ref{eq: gen dW 0}). Even this simpler method
is limited to very simple constraints and it is hard to see the underlying
universal structure and compare it the classical results. In what
follows we explore the low efficiency limit, but before doing so we
wish to point out that the solution of (\ref{eq: gen dW 0}) and (\ref{eq: G(chi,Eh)})
yields an efficiency of the form $\eta=\eta(G,\eta_{c})$. That is,
an efficiency that depends only on the constraint and on the temperature
ratio. It does not depend on the number of levels or on the engine
specific details of the energy level structure $\mc E_{h}$. Hence,
even without an explicit solution it is clear there is universality
to all order in $\chi$ for hot quantum Otto engines that are subjected
to the some constraint (or requirement). To the lowest order in $\chi$
we can expand: 
\begin{equation}
\frac{d}{d\chi}\left|\mc E_{h}\right|/\left|\mc E_{h}\right|=A+B\chi.\label{eq: d lnEh expan}
\end{equation}
Using (\ref{eq: d lnEh expan}) in (\ref{eq: gen dW 0}) lead to a
cubic equation in $\chi$. Since $\chi$ is small we use the lowest
order solution $\chi=\frac{\eta_{c}}{2}$ and replace the cubic term
by $\chi^{3}=\frac{\eta_{c}^{3}}{8}$. This yields a quadratic equation
that is correct up to order of $\eta_{c}^{3}$ . The solution is:
\begin{eqnarray}
\eta & = & \frac{1}{2}\text{\ensuremath{\eta}}_{c}+a\text{\ensuremath{\eta}}_{c}^{2}+b\text{\ensuremath{\eta}}_{c}^{3}+O(\text{\ensuremath{\eta}}_{c}^{4}),\label{eq: eta expan gen}\\
a & = & A/4,\\
b & = & B/8.
\end{eqnarray}
 In order to obtain $a$ and $b$ we need to specify a constraint
function: To evaluate $A$ and $B$ we expand (\ref{eq: G(chi,Eh)})
in powers of $\chi$. Since $G=\sum F_{k}\chi^{k}$ is constant in
$\chi$, all nonzero order multipliers $F_{i>0}$ should be zero.
In particular $F_{1}=0$ yields:
\begin{eqnarray}
a & = & \frac{1}{4}(\frac{d}{d\chi}\left|\mc E_{h}\right|/\left|\mc E_{h}\right|)|_{\chi=0}\nonumber \\
 & = & \frac{1}{4}\frac{G_{10}}{G_{10}+G_{01}}|{}_{\chi=0},\label{eq: a expression}
\end{eqnarray}
where the subscript of $G$ specify the order of derivatives with
respect to the first and second variable (the values of the variables
are omitted for brevity but they are determined by $\chi=0$: $\left|\mc E_{c}\right|=\left|\mc E_{h}\right|=\left|\mc E_{h}\right|_{\chi=0}$).
From (\ref{eq: a expression}) two important results immediately follow.
First, if the constraint is symmetric $G(\left|\mc E_{c}\right|,\left|\mc E_{h}\right|)=G(\left|\mc E_{h}\right|,\left|\mc E_{c}\right|)$,
then $G_{10}=G_{01}$ for $\chi=0$ and therefore: 
\begin{equation}
a_{sym}=\frac{1}{8}.
\end{equation}
The second result that follows from (\ref{eq: a expression}) concerns
the asymmetric case where $G_{10}\neq G_{01}$ . If $G_{10}$ and
$G_{01}$ have the same sign then:
\begin{equation}
0\le a_{sign}\le\frac{1}{4}.
\end{equation}
The two extreme values $0$ and $\frac{1}{4}$ appear in the $\left|\mc E_{h}\right|=\text{const}$
and $\left|\mc E_{c}\right|=\text{const}$ studied earlier. 

Notice that in contrast to the classical power optimization studied
in \cite{espositoEfficiencyLowDissipation}, in the quantum work optimization
studied here the functions $\eta_{\left|\mc E_{c}\right|}$ is not
necessarily an upper bound on the efficiency. For example, this is
true if the sign of $G_{\ensuremath{10}}$ is different from that
of $G_{01}$. This can be seen by comparing the leading order of the
two cases:
\begin{equation}
\eta_{quantum}=\frac{\eta_{c}}{2}+\frac{\eta_{c}^{2}}{4(1+\frac{G_{01}}{G_{10}})}+O(\eta_{c}^{3})\label{eq: eta quantum}
\end{equation}
\begin{equation}
\eta_{LD}=\frac{\eta_{c}}{2}+\frac{\eta_{c}^{2}}{4(1+\frac{\sqrt{\Sigma_{c}}}{\sqrt{\Sigma_{h}}})}+O(\eta_{c}^{3})\label{eq: eta classical}
\end{equation}
where $\Sigma_{c,h}$ are the baths relaxation time scales \cite{espositoEfficiencyLowDissipation}.
Since $\Sigma_{c}\ge0,\Sigma_{h}\ge0$ it follows that $a\le1/4$
(for $\left|\mc E_{c}\right|=\text{const}$ $a=1/4$). In contrast
in the quantum case $a$ can be larger if $G_{01}/G_{10}$ is smaller
than zero. For example, consider the constraint $\left|\mc E_{c}\right|-(1-d)\left|\mc E_{h}\right|=\text{const}$.
$G_{01}=d-1$, $G_{10}=1$ so quadratic term is $\frac{1}{4d}\eta_{c}^{2}$.
For $d=1$ we get the expected $\frac{1}{4}$ for $\left|\mc E_{c}\right|=\text{const}$,
but for smaller $d$ , $a>1/4$ . Note that $d$ should satisfy $d>\eta_{c}$
. When $d=\eta_{c}$ the Taylor series no longer converges. Physically,
beyond this point the solution is no longer an engine. A more dramatic
non-classical behavior appears when imposing the constraint $\frac{s}{\eta_{c}}\left|\mc E_{c}\right|+(1-\frac{s}{\eta_{c}})\left|\mc E_{h}\right|=\text{const}$.
This is just the $\alpha$ constraint solved before (\ref{eq: eta alpha})
with $\alpha=\frac{s}{\eta_{c}}$ . This constraint leads to a positive
definite $\left|\mc E_{h}\right|$ for any $-\infty<s<1$. This condition
also ensures that the device operates as an engine ($W>0$). Using
(\ref{eq: eta alpha}) we get: 
\begin{equation}
\eta_{s}=\frac{\eta_{c}}{2-s}.
\end{equation}
This is different from the factor of half predicted for the linear
term from classical linear response theory. In particular, the efficiency
is not bounded by the range obtained from the low dissipation theory
(\ref{eq: eta LD bound}) \cite{espositoEfficiencyLowDissipation}.
For $s=1$ the efficiency is equal to Carnot but $\left|\mc E_{h}\right|=0$
so as expected the work is zero. Notice that this constraint is not
of the form (\ref{eq: G def}) as it involves the temperatures as
well. Consequently, formulas (\ref{eq: a expression}) and (\ref{eq: eta quantum})
(as well as (\ref{eq: eta 3rd ord}) that follows) are not valid.
This can be understood by writing (\ref{eq: eta alpha}) as a series
using $\frac{1}{1-q}=\sum_{j=0}^{\infty}q^{j}$. When $s=\frac{q}{\eta_{c}}$
all powers collapse into a linear power of $\eta_{c}$. Therefore,
the truncated perturbation analysis carried out before can never give
the right result for this type of order changing constraints.

\section*{Next order for symmetric constraints}

we use the $F_{2}=0$ condition from $\text{const}=G=F_{0}+F_{1}\chi+\frac{1}{2}F_{2}\chi^{2}+O(\chi^{3})$
and get for the symmetric case ($G_{ij}=G_{ji}$):

\begin{equation}
\frac{\frac{d}{d\chi}\left|\mc E_{h}\right|}{\left|\mc E_{h}\right|}=\frac{1}{2}+\frac{1}{4}[1+\frac{\left|\mc E_{h}\right|(G_{11}-G_{20})}{G_{10}}|_{\chi=0}]\chi
\end{equation}
The multiplier of the linear term is $B$ and therefore: 
\begin{eqnarray}
\eta_{sym} & = & \frac{1}{2}\eta_{c}+\frac{1}{8}\eta_{c}^{2}\nonumber \\
 & + & \frac{1}{32}[1+\frac{\left|\mc E_{h}\right|(G_{11}-G_{20})}{G_{10}}|_{\chi=0}]\eta_{c}^{3}+O(\eta_{c}^{4})\nonumber \\
\label{eq: eta 3rd ord}
\end{eqnarray}
For example for the CA constraint $\left|\mc E_{c}\right|\left|\mc E_{h}\right|=\text{const}$,
$G_{11}=1$, $G_{20}=0$, $G_{10}=\left|\mc E_{h}\right|$ and indeed
we get the correct factor $\frac{1}{16}\eta_{c}^{3}$. As a second
example consider the efficiency $\eta_{\alpha=1/2}$ (\ref{eq: eta alpha})
obtained from the constraint $\left|\mc E_{c}\right|+\left|\mc E_{h}\right|=\text{const}$.
In this case $G_{11}=G_{20}=0$ so the multiplier of the cubic term
is $1/32$ as can be verified from the exact expression for the efficiency.
Using the same methods a similar (yet considerably more cumbersome)
formula can be written for the non-symmetric case.

\section*{Extension to colder engines }

Surprisingly the next order in $\beta$ only adds the following leading
order terms to the work:
\begin{eqnarray}
\frac{\xi}{2-\xi}\frac{1}{N}\sum_{i=1}^{N}[\frac{1}{2}\beta_{c}^{2}\mc E_{c,i}^{3}+\frac{1}{2}\beta_{h}^{2}\mc E_{h,i}^{3}\nonumber \\
-\frac{1}{2}\beta_{c}^{2}\mc E_{c,i}^{2}\mc E_{h,i}-\frac{1}{2}\beta_{h}^{2}\mc E_{h,i}^{2}\mc E_{c,i}]\label{eq: O(beta3)}
\end{eqnarray}
In principle, it complicates the optimization, however if the $\mc E_{c},\mc E_{h}$
are symmetric with respect to zero then each term individually sums
up to zero and all the results previously obtained still hold. In
particular (\ref{eq: O(beta3)}) is always zero for a two-level systems
and systems with evenly-space spectrum.

We have studied internal optimization of hot quantum Otto engines.
Universal features of the efficiency were identified. For some optimization
constraints the efficiencies at maximal work are the same as the efficiency
at maximum power in the low dissipation limit. Yet, we find constraints
for which the efficiencies deviate from the classical results, as
they cannot be obtained from perturbative analysis. In the present
case the optimization is with respect to the internal properties of
the working fluid, while in the low dissipation limit the power is
optimized with respect to the heat transport. It is interesting to
see if similar universality appears in different engines (e.g. continuous
engines) and in different operating regimes. 
\begin{acknowledgments}
Work supported by the Israel Science Foundation. Part of this work
was supported by the COST Action MP1209 'Thermodynamics in the quantum
regime'.
\end{acknowledgments}
\bibliographystyle{apsrev4-1}
\bibliography{dephc2_160614}

\end{document}